\documentclass[doublecol,linenumbers]{epl2}

\usepackage[T1]{fontenc}
\usepackage[latin9]{inputenc}
\usepackage{color}
\definecolor{note_fontcolor}{rgb}{0.80078125, 0.80078125, 0.80078125}
\usepackage{units}
\usepackage{amsmath}
\usepackage{amssymb}
\usepackage{graphicx}

 \definecolor{GREEN}{rgb}{0,0.5,0}

\makeatother

\title{Rheology of athermal amorphous solids: Revisiting simplified scenarios
and the concept of mechanical noise temperature.}
\shorttitle{Rheology of athermal amorphous solids}

\author{Alexandre Nicolas\inst{1,2}, Kirsten Martens\inst{1,2}, Jean-Louis Barrat\inst{1,2,3}}
\shortauthor{A. Nicolas \etal}

\institute{                    
  \inst{1} Univ. Grenoble Alpes, LIPhy, F-38000 Grenoble, France\\
  \inst{2} CNRS, LIPhy, F-38000 Grenoble, France\\
  \inst{3} Institut Laue-Langevin, 6 rue Jules Horowitz, BP 156, F-38042 Grenoble, France
}


\abstract{
We study the rheology of amorphous solids in the limit of negligible
thermal fluctuations. General arguments indicate that the shear-rate dependence
of the stress results from an interplay between the time scales of the macroscopic
drive and the (cascades of) local particle rearrangements. Such
rearrangements are known to induce a redistribution of the elastic
stress in the system. Although mechanical noise, \emph{i.e.}, the local stress
fluctuations arising from this redistribution, is widely believed
to activate new particle rearrangements, we provide evidence that
casts severe doubt on the analogy with thermal fluctuations: 
mechanical and thermal fluctuations lead to asymptotically different
statistics for barrier crossing. These ideas are illustrated and supported by a simple
elasto-plastic model whose ingredients are directly connected with
the physical processes relevant for the flow.
}

\begin{document}

\maketitle

A disordered assembly of interacting particles, packed densely enough for the system to be able to bear stress,
provides  a realistic image of an amorphous solid
- a heap of sand, a foam, an emulsion, a colloidal glass, a molecular
glass, etc., depending on whether the particles are grains, air bubbles,
drops, colloids or molecules. If the particles are not, or are hardly,
affected by thermal fluctuations, the material is said to be athermal.
External drive is then required to activate the dynamics of the
system. When the material is shear driven, the flow curve, that is, the relation
between the applied shear rate $\dot{\gamma}$ and the macroscopic
shear stress $\Sigma$, is often satisfactorily described by a Herschel-Bulkley equation, 
$\Sigma=\Sigma_{0}+A\dot{\gamma}^{n}$,
with an exponent $n$ usually close to, or slightly lower than, 0.5 \cite{Princen1989,Becu2006,Lemaitre2009,Schall2010,Katgert2010,Jop2012}.
An unsettled question, however, regards the connection of this dependence on the driving
\emph{velocity}
with the widely accepted picture for the \emph{spatial} organisation of the flow of disordered solids.
The slow flow scenario for these materials \cite{Argon1979,Hebraud1997,Maloney2006Amorphous,Amon2012} revolves around
localised particle rearrangements (plastic events) bursting in
an essentially elastically deforming matrix. The {}``mechanical noise'' generated by these rearrangements, \emph{i.e.}, the fluctuating stress
and strain fields that they induce in the surrounding medium, may then spark off new rearrangements, in an avalanche-like process.

In the first part of this Letter, we clarify the physical processes 
leading to the shear-rate dependence of the stress in this scenario. For all their success in capturing
various facets of the rheology \cite{Bulatov1994,Sollich1997,Picard2005,Bocquet2009,Homer2009,Dahmen2011,Vandembroucq2011}, simple
coarse-grained models in the literature generally fail to reflect these key processes consistently for athermal materials.
We remedy this deficiency by proposing a variant of such models and show that it satisfactorily reproduces the flow curve. 
In the second part of the Letter, we contemplate whether the description 
can be simplified by interpreting the fluctuating mechanical noise, which is explicitly computed in our model,
as an effective activation temperature, 
following a popular approach in another line of modelling\cite{Sollich1997}. We conclude on more general grounds that the analogy
between mechanical noise and an activation temperature is flawed.

\smallskip


Consider a dense packing of particles confined between parallel
walls and subject to a (macroscopically) constant shear rate $\dot{\gamma}$, imposed
through  successive infinitesimal displacements of one of the walls.
We start our discussion  with  an enumeration of
the time scales that subsist in the limit of vanishing shear rate. To do so, we focus
on a {}``mesoscopic'' region of the typical size of
a plastic event.
First in line comes the time scale 
\footnote{In reality, there is naturally a \emph{distribution} of such time scales. Writing, e.g., 
$\dot\gamma^{-1} \lll \tau_T $ is just a convenient way to say that values $\tau_T < \dot\gamma^{-1}$ can be neglected
in the distribution.}
 for thermally-activated structural relaxation,
$\tau_{T}$ \cite{Ikeda2012}, which diverges in the athermal limit. Secondly, the response of the
region of interest to a small displacement of the wall can take a finite time,
$\tau_{pl}$. This time will then essentially combine
the duration of a local rearrangement,
\emph{i.e.}, the time needed to dissipate
the elastic energy that was stored locally \cite{Nicolas2013b, Ikeda2012},
with the delay for shear signal transmission within
one avalanche \cite{Chattoraj2011}. $\tau_{T}$ and $\tau_{pl}$ are the only potentially relevant time scales when $\dot\gamma \rightarrow 0$.
Within a potential energy landscape (PEL) description, they
are associated with thermally-activated
hops between energy (meta)basins, and descents towards the local minimum, respectively.

The application of a \emph{finite} shear rate introduces
a new time scale, $\dot\gamma^{-1} \gamma_{y}$, which is the duration of the elastic loading phase
 prior to yield. In the PEL viewpoint, this is the {}``refresh rate'' of the PEL topology, 
owing to changes in the boundary conditions.

As stated in \cite{Maloney2006Amorphous}, quasistatic simulations, which perform
an energy minimisation after each strain increment,
rely on the following separation of the material and driving timescales: 
\begin{equation}
\tau_{pl}\lll\dot{\gamma}^{-1}\lll\tau_{T}.\label{eq:time_separation}
\end{equation}
 As long as eq.~\ref{eq:time_separation} holds, \emph{the system will follow
the very same trajectory in phase space as a function of the strain}
$\gamma\equiv\dot{\gamma} t $, regardless of the shear \emph{rate},
thereby yielding a constant elastic stress $\sigma_{el}$. One should
now recall that, for a solid-like material at low shear rate, the 
elastic stress dominates the total stress $\Sigma$ to such an extent
that the dissipative contribution to
the stress is often discarded in computer simulations, $\Sigma\approx\sigma_{el}$ \cite{Tighe2010}. Accordingly,
the only way to recover a non constant flow curve $\Sigma\left(\dot{\gamma}\right)$
involves a breakdown of the timescale separation, eq.~\ref{eq:time_separation}, and thus, for athermal materials, an
interplay between the drive and the (cascades of) localised rearrangements. In granular 
media or suspensions of hard particles, this is quantified by a dimensionless inertial or viscous number 
\cite{Boyer2011}.
More generally, the descent towards the energy minimum of the system
is disrupted by the external drive. The impossibility for the system to fully relax between strain increments
(see fig.~1(right) in ref.~\cite{Tsamados2010})
 is reflected, for instance, by the variations of the mean particle overlaps with the shear rate. 
Near the jamming transition, these variations are 
correlated to the flow curve \cite{Olsson2012}.
Deeper in the solid phase, where the simple flow scenario outlined above has
proved its consistency,  strain accumulation 
during the propagation of shear waves sets a shear-rate dependent limit on the spatial extent of the avalanches 
observed in athermal particle-based simulations \cite{Lemaitre2009}.

Surprisingly, when surveying existing coarse-grained models, one realizes that they generally do not attempt to describe the disruption of rearrangements by the drive. For instance, in  H\'ebraud-Lequeux's
model \cite{Hebraud1998}, or in the Kinetic Elastoplastic theory \cite{Bocquet2009}, as well as in Picard's model \cite{Picard2005,Martens2012},
the increase of the stress as $\dot{\gamma}$ increases derives from
the hypothesised conservation of an elastic behaviour on a given site for a constant
time (on average) \emph{after} the stress threshold has been reached locally, which appears unphysical.

Therefore, we present a variant of these models which reflects the physical processes at play.
A 2D system is discretised into linear elastic blocks of uniform shear modulus $\mu$ and of the size of a rearranging
region. To condense notations, the deviatoric stress
\footnote{Although it provides a more
 realistic description, using a tensorial description of stress,  instead of only focusing on $\sigma_{xy}$, plays virtually no role in the model. See ref.\cite{Nicolas2014u} for a discussion of these aspects.}
borne by each block ${(i,j)}$ is written $\boldsymbol\sigma(i,j)= [\sigma_{xx}(i,j),\, \sigma_{xy}(i,j)]^\top$. 
The onset of a plastic event on a given block is determined by a von Mises yield criterion: as soon as the maximal shear stress 
$\left\Vert \boldsymbol\sigma(i,j) \right\Vert \equiv \sqrt { \sigma_{xx}^2(i,j) + \sigma_{xy}^2(i,j) }$ grows larger than the local yield stress,
defined below, the block yields. One then has a stress-laden
fluid-like inclusion in an elastic medium.  An unconstrained inclusion would deform at a rate 
$\boldsymbol{\dot\epsilon}^{pl} \equiv \boldsymbol\sigma(i,j) / 2\eta_{\mathrm{eff}}$,
 with $\eta_{\mathrm{eff}}$ the effective viscosity
of the inclusion, in the overdamped regime; a time scale $\tau \equiv \eta_{\mathrm{eff}} / \mu$ 
for local energy dissipation thus arises\cite{Nicolas2013b}. However, this deformation is limited by 
the embedding elastic medium, and part of the stress borne by the inclusion is gradually redistributed to the latter.
The stress redistribution is described by an elastic propagator (matrix) $\mathcal{G}$, 
which was derived in ref.~\cite{Nicolas2013b} for a pointwise inclusion in an incompressible,
uniform, linear elastic medium, under the assumption of 
infinite shear wave velocity. As the pointwise limit of
a 2D Eshelby inclusion problem \cite{Eshelby1957}, $\mathcal{G}$ also 
features an $r^{-2}$ decay in space and a four-fold angular symmetry, in accordance
 with experimental and numerical evidence \cite{Schall2007,Puosi2014}. 

According to the above mechanism, the evolution of the local stress tensor is governed by,
\begin{equation}
\partial_{t}\boldsymbol\sigma(i,j)=\mu\boldsymbol{\dot\gamma}+2\mu \sum_{i^\prime, j^\prime}\mathcal{G}_{i-i^\prime,j-j^\prime} 
\boldsymbol{\dot\epsilon}^{pl}(i^\prime,j^\prime),
\label{eq:stress_evolution}
\end{equation}
where $\boldsymbol{\dot\epsilon}^{pl}(i^\prime,j^\prime) = \boldsymbol\sigma(i^\prime,j^\prime) / 2\mu\tau $ 
if the block is plastic, {\bf 0} otherwise. The second term on the right 
hand side of eq.~\ref{eq:stress_evolution}, describes the effect of
plastic events, \emph{i.e.}, both the nonlocal stress redistribution and the local 
stress decay. The  eigenvalues of the local component
$\mathcal{G}_{0,0}$ are of order $-0.5$, so that the stress within a plastic element 
would decay to zero on a time scale $\tau$ in the absence of external loading or elastic recovery.

To fix the distribution of yield stresses $\sigma_y$, or, equivalently, 
of energy barriers $E_y \equiv \sigma_y^2 / 4\mu$,  and the duration of a plastic event, we reason
on the basis of a schematic vision of the PEL of a rearranging region.
This landscape is composed of metabasins of exponentially distributed
depths $E_{y}$, as suggested by some experimental results on colloidal glasses \cite{Zargar2013} 
and as in the Soft Glassy Rheology (SGR) model \cite{Sollich1997}.
For practical reasons, we neglect small jumps between PEL basins and
 focus on the larger jumps between metabasins, which correspond to the irreversible
 jumps at low enough temperature \cite{Doliwa2003,Heuer2008}; to this end, we simply introduce a lower
 cut-off $E_{y}^{min}=\nicefrac{\mu\gamma_{c}^{2}}{4}$ in the energy distribution, via a Heaviside function $\Theta$, viz.,

\begin{equation}
P\left(E_{y}\right)=\Theta\left(E_{y}-E_{y}^{min}\right)\lambda e^{\lambda\left(E_{y}^{min}-E_{y}\right)},
\end{equation}
where $\lambda$ is chosen so that the average yield strain $\left\langle \gamma_{y}\right\rangle $
takes a realistic value, say, $10\%$ for emulsions \cite{Hebraud1997}. In order to describe elastic recovery, we further assume that
there is some typical distance (measured in terms of strain) between
metabasin minima. This distance is related to the parameter $\gamma_c$ used to define 
$E_{y}^{min}$; for simplicity, it is set to exactly $\gamma_c$. A block will then remain plastic until the strain
$\gamma_{c}$ has been cumulated during plasticity, that is, as long as 
$\int dt\left\Vert 2\boldsymbol{\dot\epsilon}(i,j)\left(t\right)\right\Vert <\gamma_{c}$, 
where the local rate of deformation $\boldsymbol{\dot\epsilon}(i,j)$ is the sum of the
plastic strain rate, $\boldsymbol{\dot\epsilon}^{pl}(i,j)$, and an elastic component,
$\partial_t \boldsymbol\sigma(i,j) / 2\mu$, which includes the reaction of the medium 
and the external loading (see eq.~\ref{eq:stress_evolution}).
Albeit somewhat arbitrary, this criterion for elastic recovery is physically plausible in a PEL perspective,
 and it provides a convenient way to implement the aforementioned disruption of plastic events by
the drive. At the end of the plastic event, the
local energy barrier is renewed. Apart from the time and stress units, $\tau$ and $\mu$, the 
only parameter left free in the model is the ratio
$\gamma_{c}/\langle \gamma_{y}\rangle$,
 which we set to $0.7$. (Changing this value brings on but slight variations
of the results).%

\begin{figure}
\begin{centering}
\includegraphics[width=7cm]{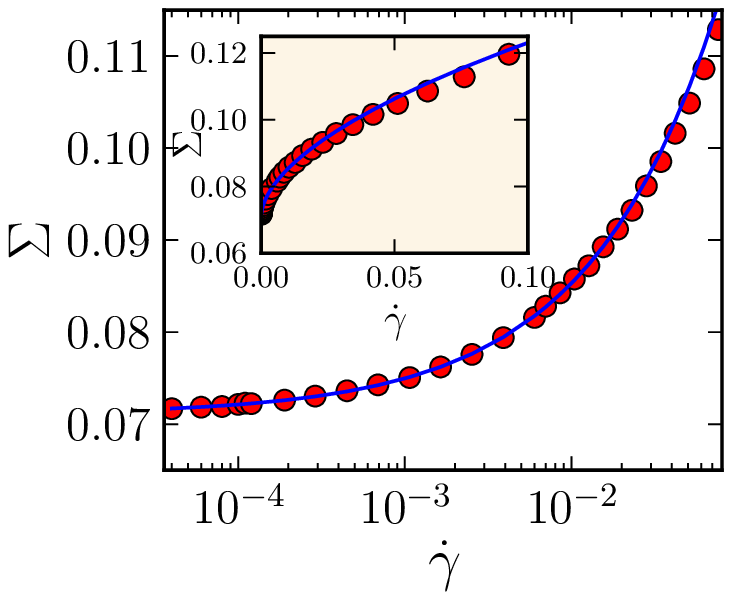}
\par\end{centering}

\caption{\label{fig:flow_curve}Macroscopic shear stress $\Sigma$ as a function
of the applied shear rate $\dot{\gamma}$, for $\nicefrac{\gamma_{c}}{\left\langle \gamma_{y}\right\rangle }=0.7.$
The solid line is the fitting curve
$\Sigma=0.071+0.19\dot{\gamma}^{0.56}$. (\emph{Inset}) Same data, in linear-linear plot.}

\end{figure}

The flow curve obtained from simulations of the model is plotted in
fig.~\ref{fig:flow_curve}. Quite interestingly, at reasonably low
shear rates $\dot{\gamma}<10^{-2}$, the curve is nicely fit by a
Herschel-Bulkley equation with exponent $n=0.56$. It is noteworthy that the Herschel-Bulkley fit
holds not only in the direct vicinity of the yield stress, as in other coarse-grained models \cite{Hebraud1997,Bocquet2009,Sollich1997}, 
but over a reasonably large window of shear rates, in 
accordance with experimental observations \cite{Princen1989,Becu2006,Schall2010}.
At higher shear rates, for $\dot{\gamma} \tau  > \left\langle \gamma_{y}\right\rangle $,
one enters a regime dominated by the dissipative stress
during plastic events, which was assumed linear 
in the strain rate here.

Note that the flow curve already rises at very low shear rates in our model. 
Around $\dot{\gamma}^{-1}\sim10^{3}$  the  life time  of a plastic event (of order $\tau$) is two orders of magnitude shorter
than $\gamma_{y} \dot{\gamma}^{-1}$, and a stress increase is already seen. This feature may be applicable to experimental systems such as foams, for which the flow curve rises even when the inverse strain rate is much longer than an elementary rearrangement of bubbles ({}``T1 event'').

Our model cannot be solved analytically,  mostly because 
of the explicit description of the mechanical noise, \emph{i.e.}, the elastic interactions mediated by $\mathcal{G}$. 
A popular class of models, with  SGR \cite{Sollich1997,Sollich1998,Fielding2000} on the frontline but also refs.~\cite{Pouliquen1996,Behringer2008,Reddy2011}, propose to identify such mechanical noise with an effective activation temperature. The flow curve is then explained in terms of activated yielding events in a random PEL: Shear lowers the associated energy barriers,
and activation is controlled by a temperature-like parameter $x$ which
presumably accounts for the mechanical noise.  At higher shear rates activation has less 
time to take place, so the system explores higher values of the stress. Let us 
first assess the validity of the activation temperature analogy within the framework of our model,
before turning to more general arguments.

To start with, notice that the model reduces to a spatially-resolved, athermal version of SGR if plastic events are made
instantaneous and allow a complete relaxation of the local stress. In this limit, varying the shear rate simply comes down to rescaling time, 
$t \rightarrow \dot\gamma t$.
The macroscopic stress is then clearly independent of the shear
rate, consistently with the then-obeyed separation of timescales:
$0=\tau_{pl}\lll\dot{\gamma}^{-1}\lll\tau_{T}=\infty$, but contrary to SGR's predictions at any $x>0$.
This is a first hint that mechanical noise is irreducible to an effective activation temperature.

\begin{figure}
\begin{centering}
\includegraphics[width=6cm]{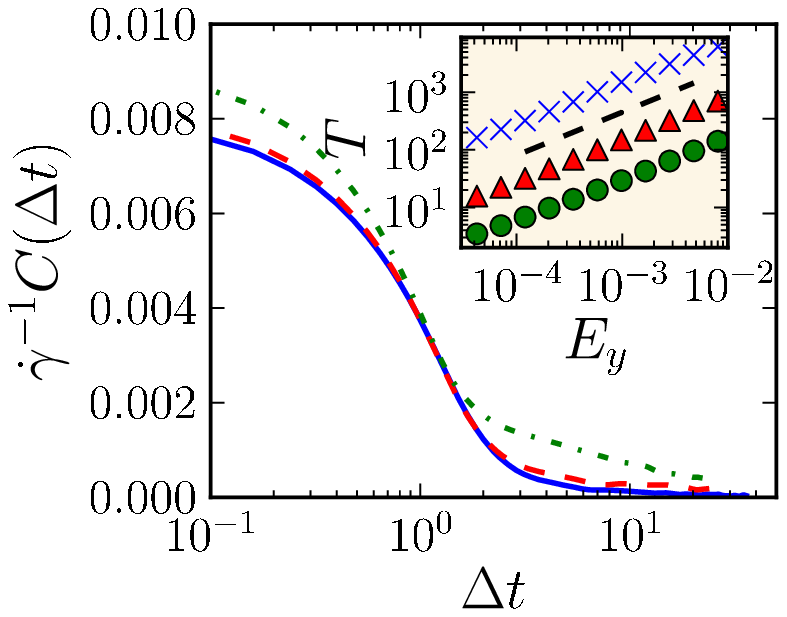}
\caption{\label{fig:mechanical_noise_hopping_time} Properties of the mechanical noise experienced by a block. Two-time autocorrelation function of its xy-component, $C\left(\Delta t\right)\equiv\left\langle\delta\dot\sigma_{xy}\left(t\right)\delta\dot\sigma_{xy}\left(t+\Delta t\right)\right\rangle $, at shear rates $\dot\gamma=$ (\emph{solid blue line}) $10^{-4}$, (\emph{dashed red line}) $10^{-3}$, and (\emph{dash-dotted green line}) $6\cdot 10^{-3}$. (Inset) Hopping time \emph{vs.} energy barrier $E_{y}$ at  $\dot\gamma=$ ($\color{blue}\times$) $10^{-4}$, ($\color{red}\blacktriangle$) $10^{-3}$, and ($\color{GREEN}\bullet$) $6\cdot 10^{-3}$. The dashed line has a slope of 0.75. System size: 256x256. Note that the use of \emph{fictitious} blocks (see text) allows to test potential barriers $E_y < E_y^{min}$.}
\end{centering}
\end{figure}

\begin{figure*}
\begin{centering}
\includegraphics[width=12cm]{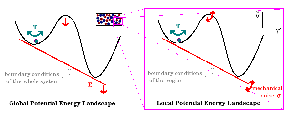}
\caption{\label{fig:PEL_sketches}Schematic representation of the difference
between thermal fluctuations and mechanical noise fluctuations in the PEL perspective.}
\end{centering}
\end{figure*}
\par

To explore the question more thoroughly,
we keep track of the mechanical noise fluctuations (per unit time) $\boldsymbol{\delta\dot\sigma}$
that a randomly selected block experiences, \emph{i.e.}, the fluctuations 
of the nonlocal terms in eq.~\ref{eq:stress_evolution}. We then study the yielding
time $T\left(E_{y}\right)$ of a fictitious block subject {\em only} to this
mechanical noise, as a function of its energy barrier $E_{y}$, by measuring how long
one has to wait before its elastic energy,
$\frac{1}{4\mu} \left\Vert \int_0^{t} \boldsymbol{\delta \dot{\sigma}}(t^\prime)dt^\prime\right\Vert^{2}$, grows
larger than $E_y$.
Note that only the mechanical noise \emph{fluctuations }can act as {}``random kicks'';
the average value, which is proportional to $\dot{\gamma}$, ought
to be treated separately, as a drift term.

The two-time autocorrelation function of the steady-state fluctuations  $\delta\dot\sigma_{xy}(i,j)$ of 
$\sigma_{xy}(i,j)$, shown in fig.~\ref{fig:mechanical_noise_hopping_time}, displays a fast
initial decay, with a decay time similar to the plastic event lifetime.
A small fraction, however, remains correlated
over much longer times, which we tentatively ascribe to long-lived correlations
in the yield stresses of nearby blocks, the latter being
renewed only every $\gamma_{y}\dot{\gamma}^{-1}$. The magnitude
of $\delta\dot\sigma_{xy}$ naturally increases with the
number of simultaneous plastic events, and therefore with $\dot\gamma$. 
Turning to the escape time $T\left(E_{y}\right)$, the
data plotted in fig.~\ref{fig:mechanical_noise_hopping_time} rule out
the Arrhenius law characteristic of activated processes,
\emph{i.e.}, 
\begin{equation}
T\left(E_{y}\right) \propto e^{\frac{E_{y}}{x}}.
\label{eq:activated_exp}
\end{equation}
 Instead, they
are in favour of a hyperdiffusive process, with a power-law scaling $T\sim E_y^{0.75} \sim \sigma_y^{1.5}$.

How general are these findings? Let us recall that, in the theory of activated processes,
a transition is completed 
when thermal fluctuations $\boldsymbol{f_\mathrm{th}}$ have pushed the system all the way up a potential barrier, in 
a \emph{fixed} PEL $V({\bf x})$ \cite{Kramers1940}. Here, ${\bf x}$ is a high-dimensional vector containing the positions of all particles.
 For concreteness, consider the Langevin equation of motion in the overdamped regime,

\begin{equation}
0 = -\zeta \frac{d { \bf x}}{dt}(t) - \nabla_{{ \bf x}} V({\bf x}) + \boldsymbol{f_{\mathrm{th}}}(t),
\label{eq:Langevin}
\end{equation}
where $\zeta$ is a friction coefficient, $\langle \boldsymbol{f_{\mathrm{th}}}(t)\rangle= {\bf 0}$, and $\langle \boldsymbol{f_{\mathrm{th}}}(t) \otimes \boldsymbol{f_{\mathrm{th}}}(t^\prime) \rangle = 2 \zeta k_B T \delta (t-t^\prime) \mathbb{I} $ (where $\mathbb{I}$ is the identity matrix). 
The exponential dependence in the Arrhenius law, eq.~\ref{eq:activated_exp}, hinges on the 
presence of recoil forces $-\nabla_{{ \bf x}} V$ that constantly oppose the uphill motion.
In contrast, mechanical noise fluctuations due to irreversible rearrangements cause \emph{persistent changes} 
to the boundary conditions of the region of interest, thereby
durably altering its PEL and stable points. 
Of course, transient effects, such as temporary dilation or inertial vibrations \cite{Salerno2012}, 
may also occur during plastic events, but, being temporary, they will be subdominant, at least for large energy barriers.

The disparity between thermal fluctuations and mechanical noise
is schematically illustrated in fig.~\ref{fig:PEL_sketches}, in which ${\bf x}$ is substituted by a 
scalar reaction coordinate, the shear strain $\gamma$. In this picture, mechanical noise 
acts as a {}``random'' external stress, which tilts the potential $V(\gamma)$ of the region (supposed of unit volume) \cite{Gagnon2001} into $\tilde{V}(\gamma, t) \equiv V(\gamma) - \gamma \sigma(t) $,
where $\sigma(t)= \langle \dot\sigma \rangle t + f_\mathrm{mec}(t)$, with the shorthand $f_\mathrm{mec}(t) \equiv \int_0^t \delta \dot\sigma (\tau) d\tau$, is the stress applied at the boundary of the region.
We examine the effect of $ f_\mathrm{mec}(t)$, \emph{i.e.}, the fluctuations around the drift term $\langle \dot\sigma \rangle t$.
Under their influence, energy barriers wax and wane, and their flattening out,
 signalling a plastic event, is therefore 
similar to a first passage time problem in a simple diffusion process over a flat landscape.  More formally, after inclusion of the mechanical noise, eq.~\ref{eq:Langevin} turns into

\begin{equation}
0 = -\zeta \frac{d\gamma}{dt}(t) -  \frac{dV}{d\gamma}[\gamma(t)] +  \langle \dot\sigma \rangle t + f_\mathrm{mec}(t) + f_\mathrm{th}(t).
\end{equation}

Mechanical noise and thermal fluctuations differ in that 
\begin{equation}
\langle f_\mathrm{th}(t) f_\mathrm{th}(t^\prime) \rangle \propto \delta (t-t^\prime),
\end{equation}
whereas 
\begin{equation}
\langle f_\mathrm{mec}(t) f_\mathrm{mec}(t^\prime) \rangle 
\ =  \int_0^t d\tau \int_0^{t^\prime} d\tau^\prime C(\tau - \tau^\prime).
\end{equation}
When the autocorrelation function $C(\Delta t)\equiv \langle \delta\dot\sigma (t) \delta\dot\sigma(t+\Delta t) \rangle$ decays quickly to zero, then $\langle f_\mathrm{mec}(t) f_\mathrm{mec}(t^\prime) \rangle \sim \mathrm{min}(t,t^\prime) $, 
and it follows that the energy barrier flattens out  under the sole influence of $f_\mathrm{mec}$, \emph{i.e.},
 $\mathrm{max}\ d\tilde{V}/d\gamma \rightarrow 0$, after a 
time $T \sim (\mathrm{max}\ dV/d\gamma)^2 \equiv \sigma_y^2$. This purely diffusive case is encountered in Picard's model (\emph{data not shown}). For the model that we introduced previously,
 the process was in fact
 hyperdiffusive, owing to the presence of slowly decaying correlations of the noise. In any case, the escape occurs much faster than in an
 activated process.

This result may seem at odds with
the numerical observation of activated processes in similar situations
in ref.~\cite{Ilg2007}. In this paper the inversion
rate of a two-state system weakly coupled to 
a simple shear flow was observed to depend exponentially on the energy barrier $V_{0}$ between
the two states, with an activation temperature $x>T_{bath}$, the bath temperature.
The apparent contradiction vanishes as soon as one notices that in the specific protocol used in ref.~\cite{Ilg2007}
the internal potential energy $V(u)$ is not durably altered by the mechanical
noise. As a matter of fact, a similar idea can be carried out within
the framework of our model\footnote{See supplemental material at http://arxiv.org/pdf/1401.6340v2.pdf, p.6-7.}
and it also yields an exponential variation
of the hopping rate with $V_{0}$.

At this stage, we must say that our conclusions concerning the (non)existence
of a mechanical noise temperature have, a priori, no bearing on some
other definitions of an effective temperature, such as those based
upon fluctuation-dissipation theorems \cite{Berthier2000, Haxton2007}, 
or the effective temperature in the Shear Transformation Zone theory, which
gives a measure of {}``disorder'' fluctuations in space, regardless
of their variations in time \cite{Bouchbinder2007}.

In conclusion, in the widely accepted scenario
for the flow of amorphous solids, a non constant flow curve in an athermal system implies
the existence of an interplay between the (cascades of) local particle
rearrangements and the driving velocity. This interplay leads to a stress increase even when a large difference exists between the
duration of a single plastic event and the inverse shear rate. The interplay mechanism differs from the widespread reliance 
on effective activation phenomena triggered by mechanical noise to explain the flow curve. 
If the onset of local yield is controlled
by a parameter akin to local stress or strain, mechanical noise fluctuations
due to distant plastic events and thermal fluctuations lead to different barrier-crossing statistics.

In the light of these findings, modifications of models such as SGR or H\'ebraud-Lequeux could be considered. 
Indeed, a simple fit of the flow curve does not establish the validity of a model or the correct assessment of the statistical properties of the mechanical noise. For a full characterisation of these properties, microscopic models or particle-resolved experiments may prove necessary.

\acknowledgments
We thank E. Ferrero for and E. Bertin for interesting discussions and detailed  comments, and  I. Procaccia and P. Sollich for their remarks  on an early version of the manuscript. 
JLB is supported  by Institut Universitaire de France and by grant ERC-2011-ADG20110209.

\end{document}


\maketitle

\section{Activation temperature measured with a two-state system}

In the main text, we have shown that the hopping statistics do not
obey the Arrhenius dependence characteristic of activated events, if
the reaction coordinate that describes a hopping event is coupled
to the mechanical noise. This is typically the case if the hop is
a plastic event that is governed by the local stress or strain, presumably.
On the other hand, an Arrhenius law is recovered when the equation of motion of
the reaction coordinate, or equivalently the potential energy of the system as a function of the reaction coordinate, 
is not durably altered by mechanical noise fluctuations.

For instance, Ilg and Barrat \cite{Ilg2007} performed molecular dynamics
simulations of the shear flow of a glassy system and introduced bead-and-spring
dumbbells in the flow. The dumbbells were maintained aligned in
the neutral direction. The inversion rate of the dumbbells was measured,
and it was shown to follow an Arrhenius law. To understand this point, one can argue that the frequent 
realignment of the dumbbells along the neutral direction erases the memory of the effect of the mechanical noise 
on the reaction coordinate, namely, the distance between the beads, thereby turning it into a nonpersistent fluctuation.

As a matter of fact, a similar attempt can be carried out in the framework
of our coarse-grained model: In every elastoplastic block, we dispose
a two-state dumbbell in the crosswise direction. The potential energy
of the dumbbell only depends on the distance $u$ between the beads
and reads, $V\left(u\right)=V_{0}\left(\frac{\left(u-u_{0}\right)^{2}-\epsilon^{2}}{\epsilon^{2}}\right)^{2}$,
with $\epsilon\ll u_{0}$. Note that the dumbbell has two ground states,
at $u=u_{0}-\epsilon$ (\emph{L}) and $u=u_{0}+\epsilon$ (\emph{R}).
During the simulation, each of the beads are advected by the velocity
field $v_{y}^{\left(ext\right)}\left(r,t\right)=\int P\left(r-r^{\prime}\right)\dot{\epsilon}^{pl}\left(r^{\prime},t\right)d^{2}r$
created by plastic events, so that,
\begin{equation}
\zeta\left(\dot{u}-\nabla v_{y}^{\left(ext\right)}\left(t\right)\cdot u_{0}\right)\simeq\frac{dV}{du}\left(u\right),
\end{equation}

where $\zeta$ is a friction coefficient. To measure the dumbbell
\emph{L}-\emph{R} inversion rate, we define the exclusive attraction
basins of the ground states \emph{L} and \emph{R} as $u-u_{0}\in]-\infty,-\delta\cdot\epsilon]$
and $u-u_{0}\in[\delta\cdot\epsilon,\infty[$, respectively with $0<\delta\approx0.5<1$
(the precise value of $\delta$ hardly affects the results).

\begin{figure}
\begin{centering}
\includegraphics[width=7cm]{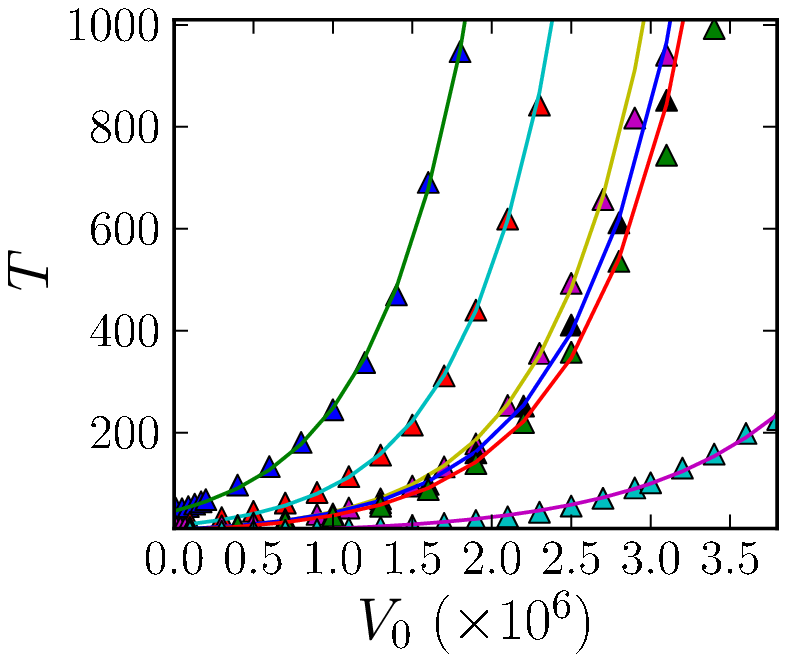}
\par\end{centering}

\caption{\label{fig:Hopping-time-Appendix}Hopping time as a function of barrier
height $V_{0}$ for $\dot{\gamma}=2\cdot10^{-4},\,5\cdot10^{-4},\,10^{-3},\,4\cdot10^{-3}$,
fitted with exponential functions $\Gamma_{0}\mathrm{exp}\left(\frac{V_{0}}{x}\right)$. }
\end{figure}

\begin{figure}
\begin{centering}
\includegraphics[width=7cm]{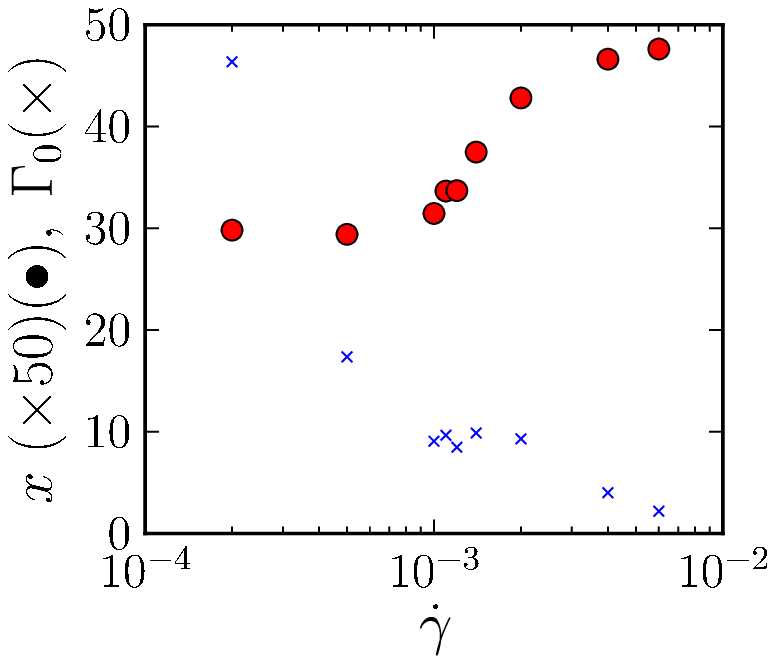}
\par\end{centering}

\caption{\label{fig:-Effective_activation_temperature}(\emph{Red dots}) Effective
activation temperature $x$ and (\emph{blue crosses}) attempt frequency
$\Gamma_{0}$ as a function of the applied shear rate $\dot{\gamma}$. }
\end{figure}

In Fig.\ref{fig:Hopping-time-Appendix}, we show the resulting hopping
times between the basins as a function of the potential $V_{0}$,
measured in units of $\zeta$, for a given $\epsilon$. Note that
the measured hopping times actually depend on $\epsilon,$ which controls
the curvature of the potential. An Arrhenius law nicely fits their
dependence on $V_{0}$, consistently with the findings of Ilg and
Barrat \cite{Ilg2007}. In Fig.\ref{fig:-Effective_activation_temperature},
the effective activation temperature associated with the Arrhenius
law is plotted as a function of the applied shear rate $\dot{\gamma}$.
As one would expect, it increases with the shear rate.

